# Narrow beam and low-sidelobe two-dimensional beam steering on thin-film lithium niobate optical phased array

*Yang Li[1,2], Shiyao Deng[1,2], Xiao Ma[1,2], Ziliang Fang[1,2], Shufeng Li[1,2], Weikang Xu[1,2], Fangheng Fu[1], Xu Ouyang[1], Yuming Wei[2], Tiefeng Yang[1], Heyuan Guan[2,\*], Huihui Lu[1,2,\*]*

[1] Guangdong Provincial Key Laboratory of Optical Fiber Sensing and Communications, Jinan University, Guangzhou 510632, China

[2] Key Laboratory of Optoelectronic Information and Sensing Technologies of Guangdong Higher Education Institutes, Jinan University, Guangzhou 510632, China

\* Correspondence: ttguanheyuan@jnu.edu.cn; thuihuilu@jnu.edu.cn

## Abstract

Optical beam steering has become indispensable in free-space optical communications, light detection and ranging (LiDAR), mapping, and projection. Optical phased array (OPA) leads this field, yet conventional versions still suffer from a narrow steering field of view (FOV), insufficient sidelobe suppression, and limited angular resolution. Thin-film lithium niobate (LN), with its strong Pockels electro-optic (EO) effect, offers a powerful integrated-photonics platform to overcome these limitations. Here we present a two-dimensional (2D) EO-steered OPA based on a non-uniformly spaced X-cut thin-film LN ridge-waveguide array. A superlattice ridge design suppresses optical crosstalk to $-20$ dB, enabling low-sidelobe far-field radiation. Using particle swarm optimization (PSO) method, we transform a uniformly spaced array into an optimized non-uniform design, largely improving angular resolution while maintaining sidelobe suppression. When combined with a single-radiating trapezoidal end-fire emitter incorporating an etched grating, the device produces a main-lobe beam width of $0.99°\times0.63°$ from an aperture of only 140 μm×250 μm, achieving a wide 2D steering range of $47°\times9.36°$ with a 20 dB sidelobe-suppression ratio. These results highlight thin-film LN OPA as a compelling route toward heterogeneous, compact, and high-performance EO beam-steering modules and ultra-miniaturized optical modulators.

## 1. Introduction

Optical modulation has become a cornerstone of modern photonics, underpinning space communications, photonic computing, and advanced imaging while continually opening new avenues for research and application. In today's era of ultra-high-speed, ultra-high-capacity data exchange, the demands on information-processing hardware have never been greater. Optical beam modulation—dynamic steering of light to a prescribed direction—now plays an indispensable role in many photonic systems.[1, 2] By changing a beam's propagation vector in free space, beam-scanning techniques provide the precise pointing needed for free-space optical links,[3] optical switching,[4] light detection and ranging (LiDAR),[5-7] and laser display technologies.[8] Robust control of beam direction thus forms the linchpin for target detection, tracking, recognition, and high-resolution imaging. High-performance photonic chips and devices constitute the core of optical beam-steering technology, enabling the transmission, reception, and processing of photon signals. LiDAR systems today employ two main classes of steering mechanisms: mechanical solutions—such as rotating mirrors and micro-electro-mechanical system (MEMS)[9]—and non-mechanical alternatives. Overall, optical beam steering can be categorized into four approaches: mechanical, acousto-optical, thermo-optical, and electro-optical (EO). Mechanical scanners are inherently bulky, require complex drive electronics, and suffer from low reliability and high fabrication costs. Moreover, their inertia limits steering speed, making them increasingly unable to meet the rapid-response demands of modern information systems.[10] Optical phased array (OPA) has emerged as a next-generation beam-steering solution,[11] offering compact footprints, high steering speeds, and precise control after decades of refinement. By independently modulating the phase of each sub-aperture, an OPA shapes the interference pattern of the composite beam, enabling agile, programmable beam direction. This capability underpins a wide—and rapidly expanding—range of applications in optical data storage and processing, free-space optical communications, and LiDAR, making OPA a major focus of international research.

Integrated photonic platforms for OPAs span silicon,[12, 13] thin-film lithium niobate (LiNbO$_3$, LN),[14, 15] III–V compound semiconductors,[16] phase-change polymers, lithium tantalate,[17] etc. Yet conventional OPA designs still face limited steering range and insufficient sidelobe suppression, motivating structural innovations and new material choices. Silicon-based OPAs leverage CMOS fabrication and low cost, but their thermo-optic tuning yields high power consumption and low speed. Thin-film LN (TFLN) waveguides, by contrast, offer ultralow propagation loss—down to 0.027 dB/cm—and a large EO coefficient ($\gamma_{33} = 30.9$ pm/V),[18] enabling efficient, fast phase modulation without thermal overhead. As semiconductor processing for TFLN matures, LN ridge-waveguide OPAs based on EO beam steering are poised to deliver wider steering angles, lower sidelobes, faster response speed, and lower power operation, positioning them as a promising platform for next-generation beam steering across communications, computing, and sensing. More recently, there have been successful LN OPAs for two-dimensional (2D) beam steering reported.[19, 20] On the other hand, the on-chip LN OPAs based on phase modulators for high-speed 2D scanning has been demonstrated.[21, 22] However, it can be noted that the TFLN OPAs reported for EO beam steering previously do not simultaneously possess efficient sidelobe suppression and narrow beam width.

Herein, we present a 16-channel thin-film lithium niobate optical phased-array

(TFLN-OPA) chip that combines a superlattice waveguide array with non-uniform widths and a trapezoidal grating emitter. A particle-swarm-optimization (PSO) algorithm refines the geometric parameters, boosting overall performance. After the fabricated chip is experimentally characterized, it delivers 20 dB sidelobe suppression and an angular resolution of 0.99°×0.63°. EO phase tuning together with wavelength sweeping further enables a wide 2D steering FOV of 47°×9.36°, underscoring the chip's potential for high-precision, high-efficiency beam control. Overall, this work demonstrates the capability of high resolution and wide-angle beam steering using such an OPA benefiting from the combination of different structures.

## 2. Device Structure and Characteristics

TFLN is an excellent platform for high-performance EO modulation thanks to its exceptional material properties. LN exhibits pronounced birefringence at an incident wavelength of 1550 nm, with an extraordinary refractive index ($n_e$) of 2.14. Figure 1a depicts the proposed TFLN OPA: a low-loss edge coupler feeds a cascade of 1×2 power splitters that evenly distribute the input light into 16 channels. To maximize the EO coefficient of LN crystal, the device is fabricated on X-cut TFLN, with the optical axis (Z-axis) lying in the chip plane and perpendicular to the waveguide propagation direction (Y-axis). Combined with a near-half-etched TFLN ridge waveguide (250 nm thick) and a silicon dioxide ($SiO_2$) cladding layer, they work together to minimize optical propagation loss. Mode-field simulations of a single ridge waveguide in the OPA yield optimal dimensions of 980 nm waveguide width ($w_{LN\_bottom}$) and 250 nm etch depth ($h_{LN\_etch}$), giving an estimated loss of just 0.015 dB/cm.

Then, etched TFLN inverted double-layer tapered waveguide and silicon oxynitride (SiON) are adopted to construct the spot size converter (SSC) as the edge coupler for efficient coupling (more parameters see Figure S1). In addition, the waveguide bending radius is set to 100 μm to reduce bending losses of OPA device. A binary tree of 1×2 multimode interferences (MMIs) connected by bend waveguides, which have a 1130 μm bending radius for achieving low excess loss and minimal beam degradation in the curved waveguides, is used for power splitting. As Figure 1b shows, the device operates with transverse electrically (TE)-polarized light, which offers stronger mode confinement than transverse magnetic (TM) polarization in this single-mode LN waveguide. Furthermore, each of the 16 channels incorporates a 5 mm-long EO phase shifter. Electrodes separated by a 4 μm gap (see Figure 1c) maximize microwave–optical overlap; because LN has a high microwave permittivity, the electric field concentrates mainly in the surrounding $SiO_2$ rather than in the ridge core. Optimizing the transmission-line structure gives electrode dimensions of 40 μm (signal), 100 μm (ground), and 1000 nm thickness, balancing low drive voltage with minimal RF loss.

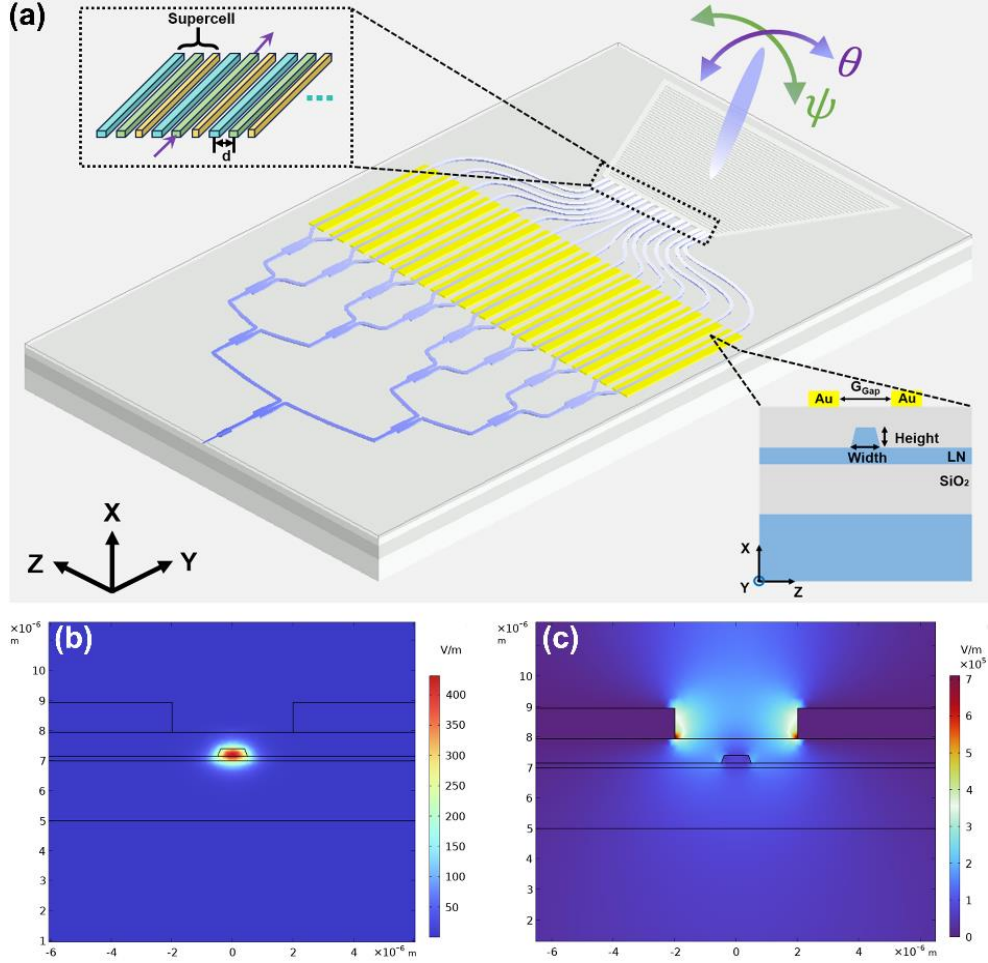

**Figure 1.** (a) Schematic diagram of the 1×16 TFLN-based OPA with non-uniformly spaced waveguides. The optical axis is aligned with the Z-axis, while the incident beam propagates along the Y-axis. Purple double arrows represent the longitudinal steering $\theta$, and the cyan double arrows indicate the lateral steering $\psi$. The top-left inset illustrates the superlattice layout in the beam-combining region, and the bottom-right inset shows the cross-section of the TFLN ridge waveguide. (b) Simulated TE-mode profile at a 1550 nm wavelength in the TFLN ridge waveguide. (c) Simulated electric-field distribution when a 1 V bias is applied to the LN waveguide electrodes.

To more effectively adjust the propagation mode of beam, the waveguides in the beam-combining region are patterned into a periodic superlattice whose widths are precisely engineered. Meanwhile, this deliberate modulation of structural parameters provides additional control over light propagation while the alternating-width pattern markedly suppresses crosstalk between adjacent channels.[23] As illustrated in the upper-left inset of Figure 1a, the beam-combining section employs a periodic superlattice in which every supercell (SC) comprises three waveguides of distinct widths $w_1 > w_2 > w_3$. All channels share a uniform center-to-center pitch $d$, and light propagates in the direction marked by the purple arrow. Fabrication limits dictate the upper and lower bounds of the waveguide width, within which the three values are judiciously spaced. Adopting a monotonically descending width sequence inside each supercell minimizes phase mismatch among neighboring guides and thereby suppresses optical crosstalk, which refers to the normalized power coupled from one waveguide into an adjacent one.[24, 25]

The proposed waveguide superlattice can utilize the difference in the width of adjacent waveguides to minimize the coupling between array elements, thereby effectively reducing the optical crosstalk in the combining region of the OPA. The waveguide width used in the OPA based on uniform spacing distribution is 980 nm. Taking the average waveguide width available in the preparation process as the reference, the superlattices with different waveguide widths of *n* are obtained by alternating arrangement. The far-field radiation intensity spectra shown in Figure 2 is obtained for comparative analysis.

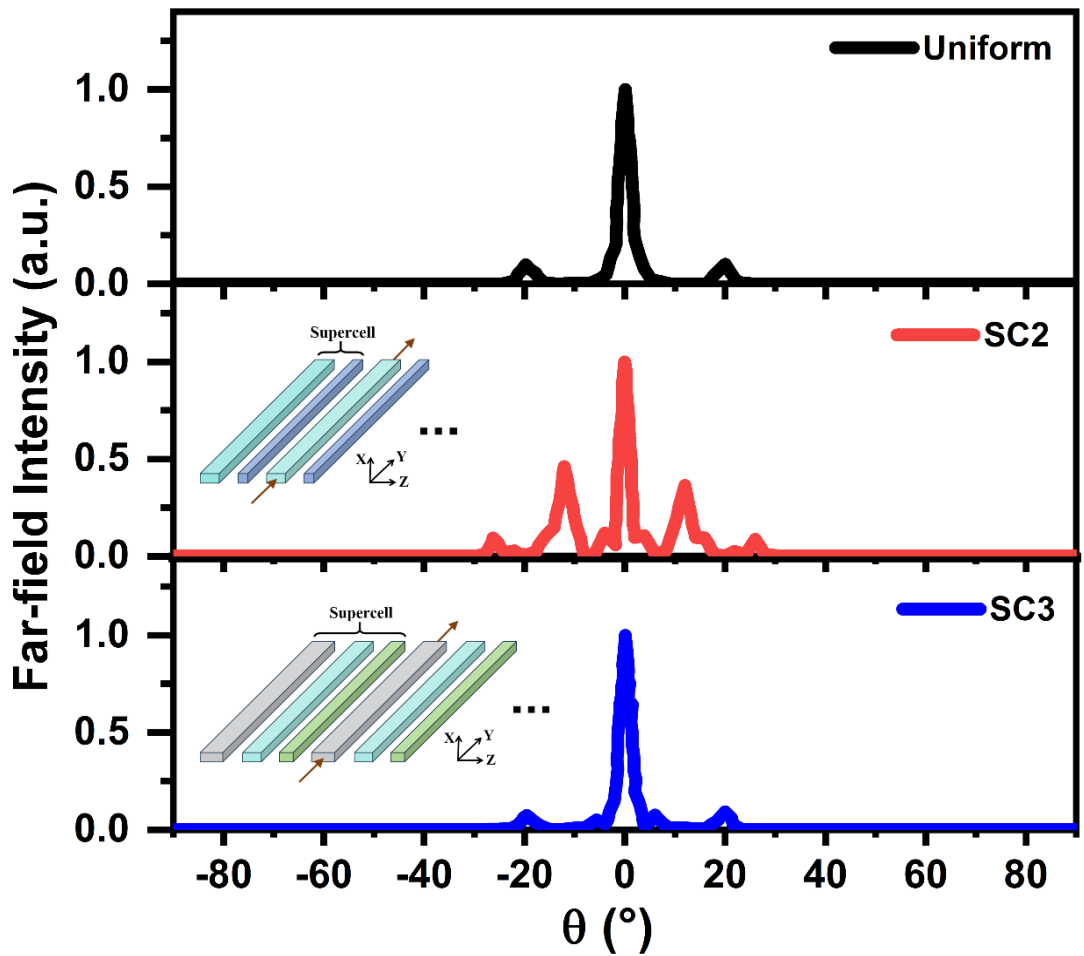

**Figure 2.** Numerical simulations of the 1×16 TFLN-based OPA reveal how the waveguide-array geometry shapes the far-field beam intensity profile. The output beam of three designs are compared: the uniform array, with the waveguide width of all being 980 nm; superlattice waveguide SC2, whose waveguide width periods are 980 nm and 860 nm, respectively; superlattice waveguide SC3, whose waveguide width periods are 1040 nm, 980 nm and 920 nm, respectively. Insets in each spectrum sketch the corresponding width patterns, with color coding denoting the different waveguide widths.

For comparison, Figure 2 compares the resulting far-field radiation spectra obtained from 1×16 TFLN OPA based on different waveguide width periods, namely:

uniform waveguide width array, with waveguide widths of 980 nm for all; superlattice waveguide SC2, whose waveguide widths alternate between 980 nm and 860 nm; superlattice waveguide SC3, whose waveguide widths follow a three-element cycle of 1040 nm, 980 nm, and 920 nm. Insets show the corresponding waveguide width-period schematics. To the best of our knowledge, the highest-intensity beam is designated as the main lobe in each far-field pattern, while subsidiary peaks are side lobes. The peak side lobe level (PSLL) quantifies the strongest side lobe relative to the main lobe and is expressed in decibels (dB). Based on this definition, the PSLL can be calculated as:[26]

$$\text{PSLL} = 10\log_{10}\left(\frac{I_{Max\ sidelobe}}{I_{Mainlobe}}\right)\ (dB) \qquad (1)$$

Analysis shows that the SC3 superlattice design drives the PSLL down to roughly −10 dB while keeping overall output-power uniformity of the array within about −20 dB.

To concentrate energy in the main lobe and suppress superlattice side lobes, we employ an appropriate PSO algorithm[27, 28] to fine-tune the spacing between waveguide elements (see Figure S2). The fitness function weighs two metrics from the array's simulated far-field pattern: the PSLL and the main-lobe width, quantified by the full width at half maximum (FWHM). By iteratively updating the inter-element spacing, the PSO converges on a configuration that simultaneously minimizes PSLL and shrinks the FWHM. According to the FWHM expression for the far-field pattern:[11]

$$\theta_{FWHM} = \frac{0.886\lambda}{Nd\cos\theta_{Beam\ Steering}} \qquad (2)$$

Where $N$, $d$ and $\theta_{Beam\ Steering}$ represent the number of array elements, element spacing and steering angle, respectively. Thus, the effective way to improve the trade-off between beam steering range and beam resolution is to increase the number of elements. However, the increase in the number of arrays will bring additional optical loss and fabrication difficulty. Therefore, the 1×16 OPA device obtained by PSO optimization can maintain a smaller number of arrays while considering high-resolution beam steering.

It is noteworthy to discuss the emission region of the OPA. In our case, we choose the end-fire trapezoidal plate length $L_{Trapezoid}$ as 350 μm according to the process requirements. To narrow the main lobe width of the far-field output beam, further simulation and binary optimization are carried out on the top and bottom widths of the emitting trapezoidal plate, as shown in Figures 3a-3b. In addition, the flat grating is designed to be trapezoidal in shape, and a suitable tilt angle needs to be selected to make it larger than the in-plane divergence angle of the flat mode, so that more than 90% of the beam power can be transmitted forward freely without being affected by the reflection of the side walls[21]. The optimal trapezoidal plate width selected needs to comprehensively consider its corresponding far-field sidelobe suppression ratio (as shown in Figure 3a) and the far-field beam main lobe width (as shown in Figure 3b). After comprehensive consideration, the optimal values are $W_{Trapezoid\_short}$=140 μm and $W_{Trapezoid\_long}$=250 μm.

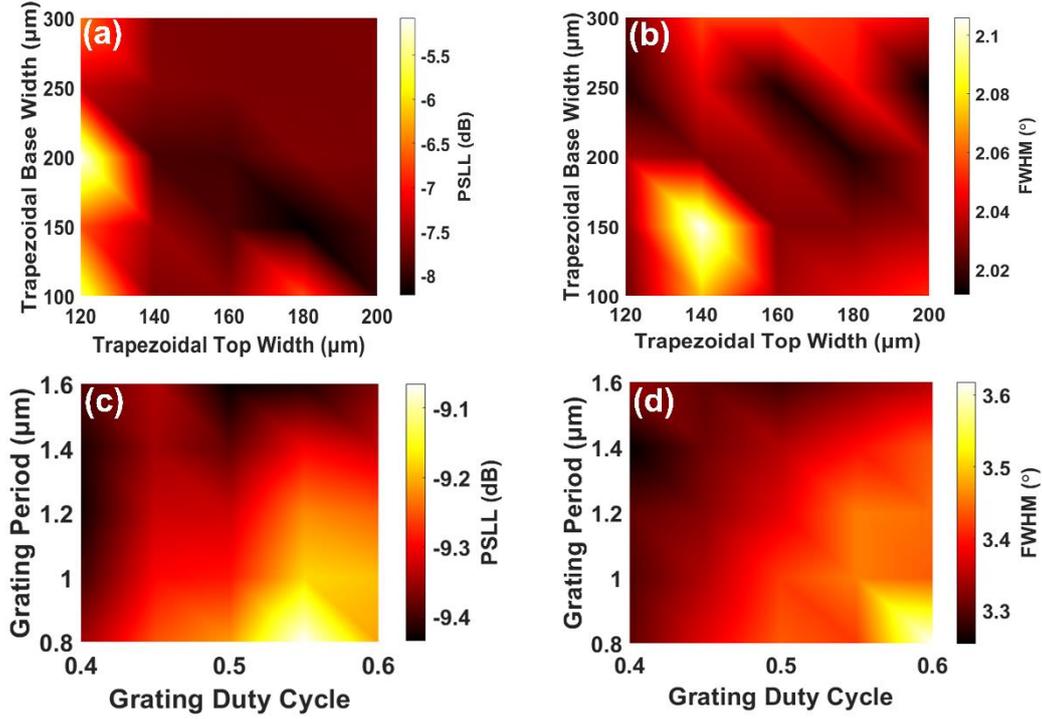

**Figure 3.** Numerical analysis of the trapezoidal transmitting grating structure in OPA. (a) When the length of the trapezoidal plate is 350 μm, the result of PSLL obtained by optimizing the upper and lower widths of the trapezoidal plate. (b) Optimization results of FWHM obtained for the upper and lower widths of the trapezoidal plate. (c) Simulation optimization results of PSLL obtained by etching grating period and its duty cycle. (d) Simulation optimization results of FWHM obtained by etching grating period and its duty cycle.

Earlier research work has shown that etching a periodic groove structure on the emitting array of a LN waveguide array to form a grating array antenna is an effective solution to improve the quality of its output beam, which helps to achieve the shaping and steering of narrower steering beam. Therefore, the etched grating on a trapezoidal plate as the end-fire of OPA is proposed, rather than using multiple emitting arrays such as a grating antenna array. When the optical waveguide mode propagates to the trapezoidal grating, the waveguide mode will be transformed into a plate divergent mode due to the in-plane interference steering combined with the out-of-plane radiation of the etched grating. Moreover, in this work, the optimal value of the etched grating thickness of 140 nm is obtained by almost half-etching the thickness of the LN ridge waveguide. The width of the etched grating used here is of the gradient type, and it will be reasonably designed according to the structural parameters of the end-fire trapezoidal plate. In order to obtain the optimal output beam steering performance of the trapezoidal plate etched grating, we conduct a binary optimization analysis on the period and duty cycle of the etched grating based on the performance indicators of the steering beam, such as the sidelobe suppression and its far-field main lobe beam width. The optimization range is the etched grating period of 800 nm~1600 nm, and the grating duty cycle of 0.4~0.6. Figures 3c-3d respectively show the binary optimization results of PSLL and FWHM obtained by etching the grating period and its duty cycle. It can be analyzed that the structural etching depth of the optimized width-gradient silicon dioxide grating in this work is 140 nm, and the etching period length of the grating is 500 nm, corresponding to a duty cycle of 0.5.

Based on the proposed 16-channel TFLN-OPA by EO steering, high-precision processing technology is used to complete chip preparation and experimental characterization for subsequent experimental research. The fabrication of the OPA chip by hard mask etching (more details can be found in Figure S3) is mainly accomplished through technologies such as electron beam lithography (EBL), inductive coupled plasma (ICP) etching, and plasma-enhanced chemical vapor deposition (PECVD). Based on this scheme, the OPA chip is fabricated on a commercial X-cut TFLN wafer, which consists of a 400 nm-thick LN layer, a 2 μm silicon dioxide layer, and a 525 μm silicon substrate layer. This multilayer structure design ensures good optical confinement effect and mechanical stability of device. The top width of the LN ridge waveguide is 980 nm, which is optimized for better single-mode transmission. The etching depth of the LN waveguide layer is 250 nm, ensuring good optical mode confinement. On the other hand, the silicon dioxide coating is deposited using PECVD technology to improve the mode field distribution and provide protection. The fabricated TFLN-OPA chip is visually presented through optical microscopy characterization, as shown in Figure 4a, which is the microscopic image of the well-prepared OPA chip for realizing 2D beam steering.

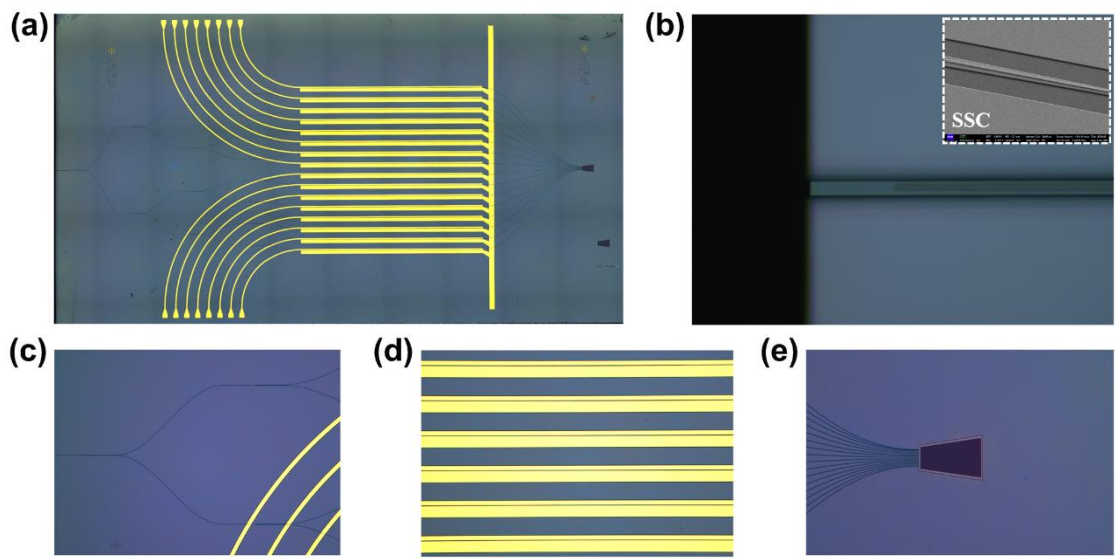

**Figure 4.** Optical micrographs of the fabricated OPA chip and its basic components. (a) Optical microscopy of the non-uniform TFLN-OPA chip. (b) Designed spot size converter (SSC) at the incident end of the chip, with its scanning electron microscopy (SEM) image in the inset. (c) Microscope of MMIs. (d) Micrograph of a portion of the optical phase shifter and the electrodes at its two ends. (e) Micrograph of the beam combining region and the transmitting end of the phased array, including superlattice waveguide array and trapezoidal emission grating.

Figures 4b-4e are electron microscope images showing the basic components of the magnified OPA chip, including the zoomed-in SSC, the corresponding MMIs based on LN ridge waveguide, the optical phase shifters of the chip and the electrode structure

distributed at both ends, and the beam combining region and the transmitting end of the phased array. It includes the microscopic diagram of the superlattice waveguide array and trapezoidal emission grating. The fabricated 16-channel OPA with an appropriate sparse non-uniform antenna spacing can be used to suppress the grating lobes and higher resolution beam steering.

## 3. Experimental Results and Discussions

In general, higher resolution can be obtained by compressing the optical beam divergence, which requires a larger aperture. According to the far-field theory $(n_{eff}A^2/(D\lambda) < 1)$,[29, 30] where A is the aperture size of the OPA, λ is the operating wavelength, D is the distance of the observation point from the array, and $n_{eff}$ is the effective refractive index of propagation inside the LN ridged waveguide. The proposed OPA based on chip-scale TFLN platform can achieve 2D beam steering. To show this capability, the small optical aperture of 140 μm×250 μm is designed to obtain high-resolution beam steering. The experimental setup for OPA is schematically shown in Figure 5a. To characterize the optical performance of the fabricated OPA chip, we have used a continuous-wave tunable laser source (AQ2200-136) to generate light ranging from 1440 nm to 1640 nm. Moreover, it is necessary to connect special optical fiber (Ultra-High Numerical Aperture 7, namely UHNA-7) to achieve efficient edge coupling between the optical fiber and the device under test (DUT). As noted, the incident laser is set to the fundamental transverse electric ($TE_0$) mode, which can be controlled by the polarization controller (PC). The far-field pattern distribution is captured by a charge coupled device (CCD). A CCD camera captures top-down images of the scattered light at the designed TFLN OPA. The emission from the OPA chip is collected by an objective lens with a numerical aperture (NA) of 0.4 and expanded through a 4-f system to fit the CCD camera.

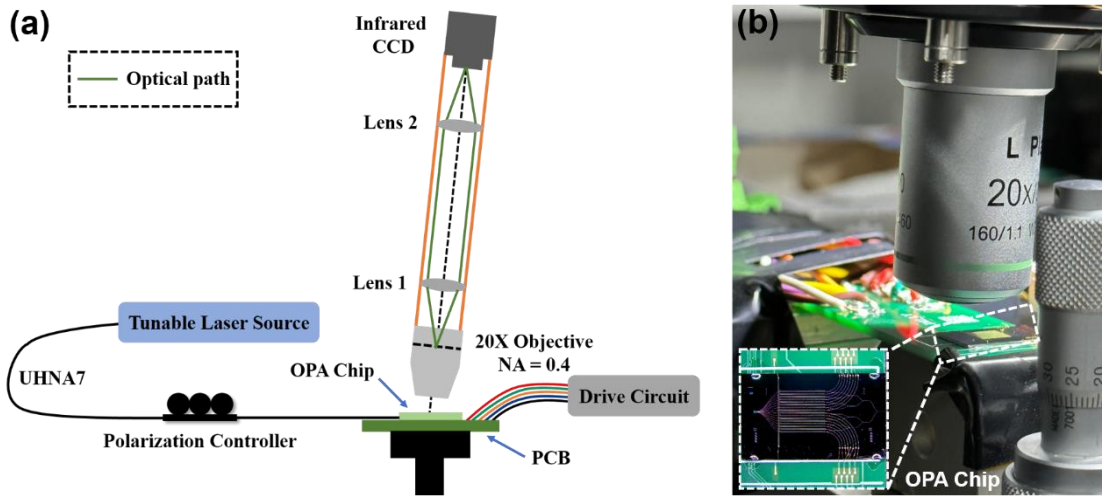

**Figure 5.** Experimental setup of the device under test (DUT). (a) Schematic diagram of the experimental setup for near-field and far-field testing of TFLN-OPA chip. (b) Photograph of the part of the experimental setup. The OPA chip is mounted on a printed circuit board (PCB).

As shown in Figure 5b, the 16-channel TFLN OPA is fabricated on a 2.1 cm×2 cm X-cut lithium niobate on insulator (LNOI) chip, which is wire-bonded on a printed circuit board (PCB) and controlled by the circuit control system that has been built. In practice, due to the inevitable technological errors in the chip processing, it is difficult to ensure that the initial phase of each radiation beam remains the same as theoretical design value. Therefore, it is necessary to adopt the phase optimization calibration method to achieve precise pointing of the radiation beam. As presented in Figure 6a, When no voltage is applied to the OPA , multiple scattered light spots of varying brightness can be clearly seen in the initial far-field interference pattern obtained. At this time, the main lobe and side lobe of the far-field beam cannot be distinguished. This is due to the initial phase difference between adjacent waveguides of the phase shifter. The EO phase is randomly distributed, resulting in the dispersion of the beam energy. To address this issue, it is necessary to perform initial phase calibration on the chip. Then, by optimizing and adjusting each voltage applied to the phase shifters of OPA, the emitted light spot is optimized to the brightest single-mode far-field spot as shown in Figure 6b. Thus, the calibration of the initial phase error of the chip is completed. At an operating wavelength of 1550 nm, the far-field beam has a FWHM spot size of 0.99°×0.63°.

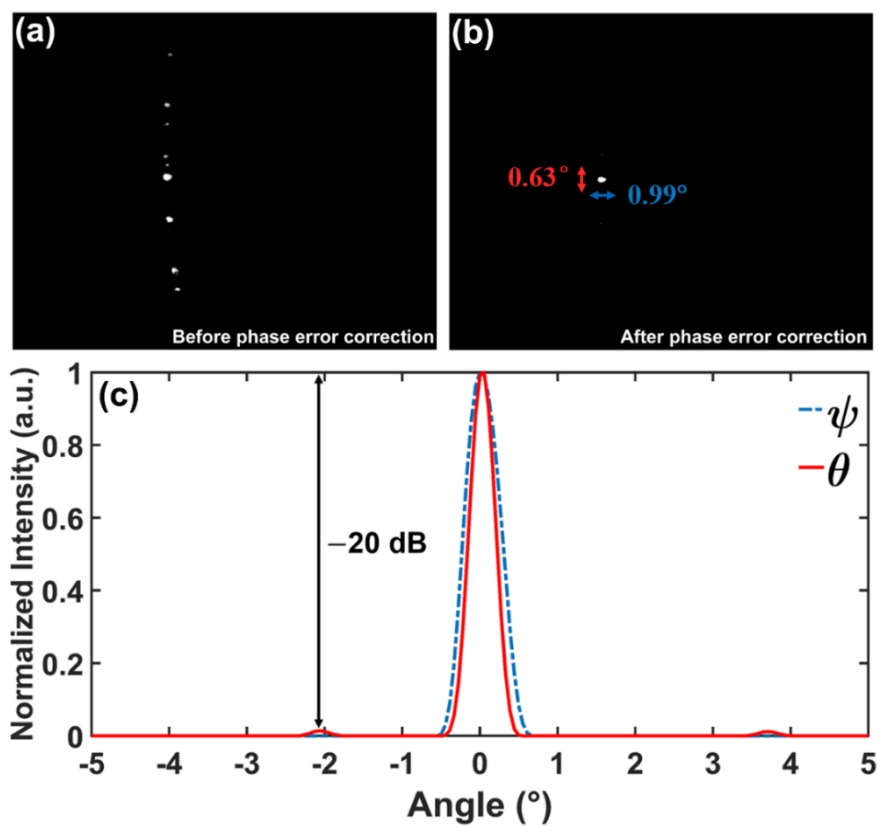

**Figure 6.** Far-field radiation spot before and after initial phase error calibration at optical wavelength of 1550 nm. (a) Far-field pattern before phase error calibration. (b) Far-field beam pattern after phase error calibration, with a well-focused main beam, showing the main-lobe width of 0.99°×0.63°. (c) Normalized far-field radiation intensity spectrum, presenting the optimized sidelobe suppression up to −20 dB.

Moreover, it can be known from the normalized beam spectra obtained in the longitudinal and lateral directions of the far-field radiation as plotted in Figure 6c, it can be seen that the calculated PSLL can be as low as −20 dB, which further verifies the necessity of completing the phase error calibration before the experimental test. Further as mentioned above, the experimental research on the beam steering of the chip is continued.

In this work, the incident laser is edge-coupled into the waveguide and modulated by 16-channel EO phase shifters integrated on the TFLN-OPA chip. Figure 7a illustrates longitudinal beam steering in the far field is achieved by applying optimized voltages to the 5-mm-long modulation region of OPA chip. Then the steering angle $\theta$ of the EO modulated light beam is theoretically given as:[31]

$$\sin\theta = \frac{\lambda \Delta \varphi}{2\pi d} \tag{3}$$

Here, $\Delta\varphi$ refers to the phase difference between the waveguide arrays. $d$ represents the array spacing of the OPA, and the operating wavelength is denoted by $\lambda$. As is evident, the side lobes of the measured far-field pattern are −13 dB, which can be tuned to approximately 47°, corresponding to a π-phase shift of the aperiodic antenna pattern, as presented in Figure 7b.

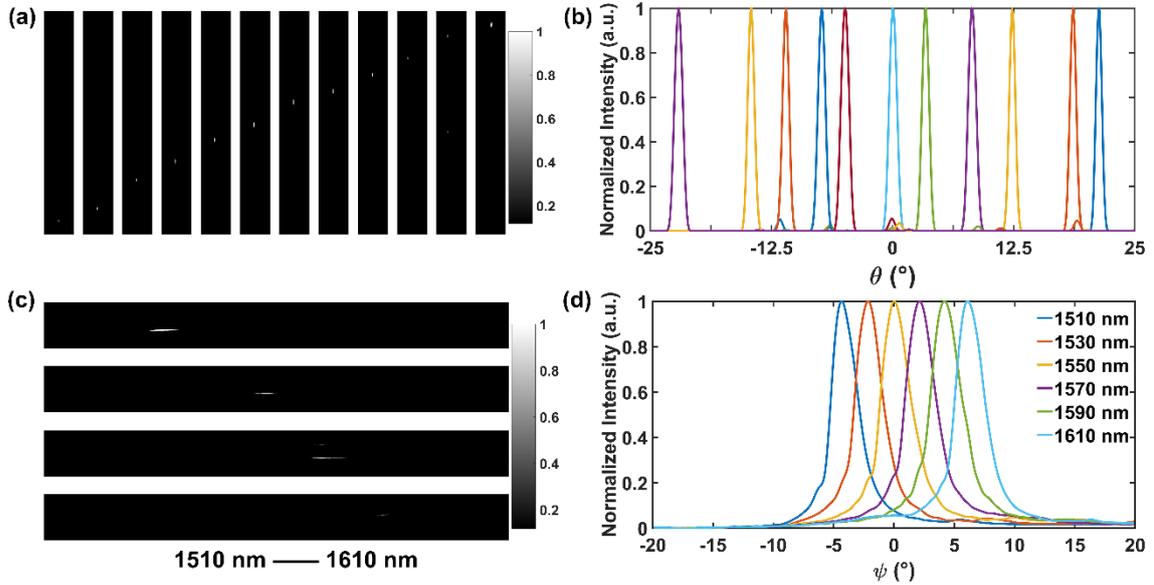

**Figure 7.** 2D beam scanning of TFLN-OPA chip. (a) Far-field radiation pattern at different steering angles obtained after applying an external voltage for phase modulation with a fixed incident wavelength of 1550 nm. (b) Measured relationship between the output power distribution and the EO modulation angle. As the applied voltage changes, the steering FOV is 47°, and the corresponding PSLL is −13 dB. (c) Far-field radiation distribution obtained by changing the wavelength of the incident laser from 1510 nm to 1610 nm. (d) Measured relationship between the output optical power distribution and the wavelength tuning angle. As the wavelength changes, the steering FOV is 9.36° and its PSLL is −12 dB.

Additionally, the RC-limited modulation bandwidth can be calculated up to 0.96 GHz with an estimated capacitance of 2.5 pF. The electrical energy for inducing a phase shift of π can indicate the low power consumption of 42.2 pJ/π per EO modulator. The grating emitter of OPA emits light. Based on the grating diffraction theory, the OPA chip can perform wavelength-tuned beam steering by changing the incident light wavelength (near-field experiment can be seen in Figure S4). The steering of the beam in the lateral dimension direction achieved by tuning the wavelength, with the output coupling angle $\psi$ controlled by the grating equation as follows:[32]

$$\sin\psi = \frac{\Lambda n_{eff} - \lambda}{n_0 \Lambda} \tag{4}$$

where $\Lambda$ is the grating period, $\lambda$ is the laser wavelength, $n_{eff}$ is the effective index of the waveguide mode, and $n_0$ is the index of the background. In the experiment, wavelength-dependent measurements are performed to help validate second-dimensional beam steering, as demonstrated in Figure 7c. By sweeping the input wavelength range from 1510 nm to 1610 nm, the steering range of the emitted light in the $\psi$-direction is 9.36°, with PSLL of −12 dB, as shown in Figure 7d.

Based on the above analysis, the proposed TFLN-OPA chip can achieve a wide beam steering range of 47°×9.36° with low sidelobe in the phased-array steering axis and the wavelength-tuning steering axis. In addition, multiple output far-field radiation modes can be obtained through wavelength tuning and phase EO modulation to superimpose and synthesize arbitrary images, such as the 2D beam trajectory EO modulated writing of the capital letter "JNU" shown in Figure 8.

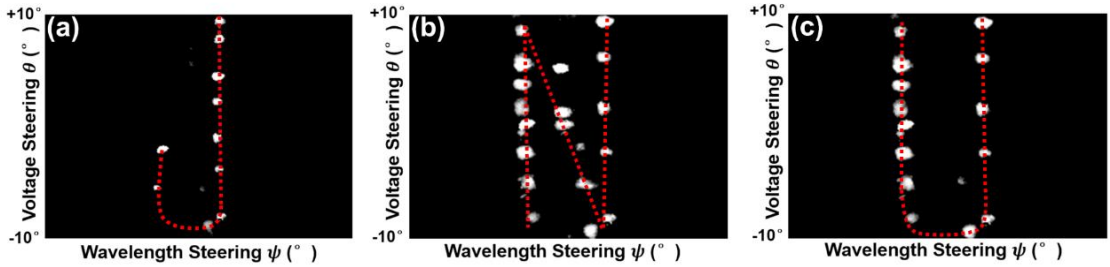

**Figure 8.** By simultaneously controlling the wavelength tuning and the phase modulation of the OPA chip, the far-field radiation pattern obtained within the steering FOV is superimposed to synthesize the 2D images of the three letters "JNU", which are (a) "J"; (b) "N"; (b) "U"; Among them, the obtained far-field beam composite images are all depicted and fitted with red dotted lines for viewing.

As a result, the OPA chip can scan the high-resolution optical beam with side lobes suppression. Notably, the steering range will be limited by the NA of objective lens, which has an acceptance angle of 48° yields. To maximize the observable FOV, devices such as rotating disks can be added to expand the far-field radiation steering range of the OPA device. Table 1 shows recent work on different kinds of OPAs compared with our work. By comparing and analyzing the optical beam steering performance of other OPAs, the TFLN-OPA chip in this work has a relatively low PSLL, and can achieve a larger steering FOV while maintaining a higher resolution. Its modulation speed is relatively fast. The results show that the proposed TFLN OPA obtain a low PSLL and finer beam with relative wide-FOV optical beam steering for only 16 channel waveguides.

Table 1. Performance summary and comparison of the reported OPAs.

| Process technology (Channels) | Modulation Mechanism | PSLL (dB) | Steering FOV | FWHM | Bandwidth | Power Consumption |
|---|---|---|---|---|---|---|
| Si OPA[33] (1×64) | Charge-injection | −8.8 | 8.9°×2.2° | 0.92°×0.32° | 324 MHz | 17 mW/π |
| Si OPA[34] (1×64) | Carrier depletion | −10 | 36°×3.4° | 0.85°×0.25° | 700 MHz | 0.45 nW/π |
| Polymer waveguides[35] (1×8) | EO effect | N/A | 19.1° | 3.9° | 2 MHz | 47.5 μW/π |
| InP OPA[36] (1×8) | EO effect | −11.7±1.8 | 18° | 2.48°±0.3° | N/A | N/A |
| LN OPA[19] (1×16) | EO effect | −10 | 24°×8° | 2°×0.6° | 4.2 GHz | 13.5 pJ/π |
| TFLN OPA[21] (1×16) | EO effect | −5.3 | 50°×8.6° | 0.73°×2.8° | 2.5 GHz | 330 pJ/π |
| TFLN OPA[22] (1×32/48) | EO effect | −7.4 | 62.2°×8.8° 40°×8.8° | 2.4°×1.2° 0.33°×1.8° | 69 MHz | 1.11 nJ/π |
| TFLN-SiN OPA[20] (1×16) | EO effect | N/A | 22°×5° | 1.2°×0.2° | 1.2 GHz | 3.2 pJ/π |
| **TFLN OPA (1×16) (This work)** | **EO effect** | **−20** | **47°×9.36°** | **0.99°×0.63°** | **0.96 GHz** | **42.2 pJ/π** |

## 4. Conclusion

In conclusion, an effective solution for suppressing sidelobes in the TFLN-based OPA with high-resolution 2D beam steering has been proposed and experimentally demonstrated. The optimization of the waveguide array geometry and validating beam-steering performance through simulation and experiment. First, superlattice width is introduced to the waveguide array, significantly lowering optical crosstalk and reducing sidelobe levels. Next, a PSO routine converts the uniformly spaced array into a non-uniform configuration, balancing sidelobe suppression with enhanced steering resolution. In addition, the trapezoidal grating emitter of the chip is systematically simulated and analyzed to achieve more efficient far-field radiation. The fabricated 1×16 TFLN OPA, featuring optimized non-uniform emitter spacing and a compact 140 μm×250 μm aperture, delivers a main-beam resolution of approximately 0.99°×0.63°.

EO phase tuning and wavelength sweeping enable a wide steering FOV of 47°×9.36°, while sidelobe levels are held 20 dB below the main lobe. These results confirm the excellent EO beam steering performance of the OPA, highlighting TFLN as a promising platform for ultra-compact, high-performance optical beam modulation devices.

## Supporting Information

Supporting Information is available from the Wiley Online Library or from the author.

## Acknowledgments

This research was funded in part by the National Key Research and Development Program of China (2023YFA1407200), the NSAF (U2330113, U2230111, U2030103), the National Natural Science Foundation of China (62405114, 12404577), the Guangdong Basic and Applied Basic Research Foundation (2025B1515020096, 2023A0505050159, 2022A1515110970, 2023A1515110626), the Fundamental and Application Foundation Project of Guangzhou (2024A04J3974).

## Conflict of interest

The authors declare no conflict of interest.

## Data Availability Statement

The data that support the findings of this study are available from the corresponding author upon reasonable request.